\documentclass[ amsmath,amssymb, aps, onecolumn, 10pt, a4paper, pra]{revtex4-2}
\usepackage{graphicx} 
\usepackage{amsfonts, amsthm}
\usepackage{lmodern}
\usepackage[T1]{fontenc}
\usepackage{mathrsfs}
\usepackage{hyperref}
\usepackage{orcidlink} 

\begin{document}

\title{Bondi Flow from Various Perspectives}
\author{Souvik Ghose\orcidlink{0000-0002-2475-0186}}
\email{dr.souvikghose@gmail.com}
\affiliation{Harish-Chandra Research Institute, HBNI, Chhatnag Road, Jhunsi, Prayagraj,  211019, India}

\author{Tapas K. Das\orcidlink{0000-0002-7013-0383}}
\email{tapas@hri.res.in}
\affiliation{Harish-Chandra Research Institute, HBNI, Chhatnag Road, Jhunsi, Prayagraj, 211019, India}

\begin{abstract}
\noindent
    Realization of the stationary integral solutions of steady state transonic accretion flow in spherical symmetry (so called Bondi flow) is crucial, since it helps to understand accretion phenomena on various astrophysical objects. Such flows are studied in the literature, in general, only from the astrophysical contexts. In recent years, however, attempts have been made to study accreting black hole systems as an example of autonomous dynamical systems. The fixed point solution scheme has been borrowed from the theory of dynamical systems to study the transonic properties of accretion flow onto an astrophysical black hole, and it has been demonstrated that the nature of the phase orbit for the transonic flow solutions can be understood even without constructing the integral solutions. Since  a large scale astrophysical fluid flow is vulnerable to external perturbation, one needs to ensure that the stationary accretion solutions are stable under perturbation. Such a task can be accomplished by adopting a  time dependent stability analysis scheme for the accretion flow, to demonstrate, under which condition the perturbation will not diverge. While performing such stability analysis, it has also been observed that a space time metric, dubbed as the sonic (or acoustic) metric can be constructed to describe the propagation of the perturbation embedded within the accreting fluid. Such a sonic metric is similar in nature to certain representation of the Schwarzschild metric. An acoustic metric, thus, mimics a black hole like spacetime within the accreting fluid, and the transonic surface can be identified with a black hole like horizon. Such identification is accomplished using the theory of causal structure, by constructing Carter-Penrose diagrams.  An accreting black hole system, thus, can be perceived as an classical analogue gravity model naturally found in the universe. Hence, accretion phenomena onto astrophysical black holes can be looked upon from three apparently non overlapping perspectives - astrophysical processes, theory of dynamical systems, and emergent gravity (alternatively, the analogue gravity) phenomena, respectively. The present article illustrates, by taking the simplest possible accretion flow model,  how one can study astrophysical accretion processes from three aforementioned perspectives.
\end{abstract}
\maketitle

\section{Introduction}
Accretion refers to the process by which matter through its gravitational attraction. Accretion is ubiquitous in astrophysics and is responsible for the formation of most astronomical objects, including galaxies, stars, planets {and compact astrophysical objects}. 
The importance of accretion of interstellar gas has been recognized since the late 1930s, primarily through the pioneering work of Hoyle, Lyttleton, and Bondi \citep{hoyle_1,hoyle_2,hoyle_3,hoyle_bondi}. However, these early studies overlooked the impact of pressure in the in-falling matter, assuming that any generated heat would be rapidly radiated away and that the gas temperature remained low, thereby justifying the neglect of pressure effects compared to dynamical effects.
In 1952, Bondi's seminal paper \cite{bondi1952spherically} marked a significant advancement {in this field} by exploring gas accretion in an adiabatic scenario, incorporating pressure effects for the first time. He considered a simplified model where an accretor of mass \(M\) was placed in an infinite, gas cloud with uniform density \(\rho_{\infty}\) and pressure \(p_{\infty}\) at an infinitely large distance from the accretor. The accretion of gas was assumed to be spherically symmetric and steady, with the mass of the accretor remaining constant during the accretion process.

In the context of black holes, accretion processes are particularly critical. Black hole accretion involves the accumulation of surrounding {compressible astrophysical fluid} into an accretion disk, where intense gravitational forces heat the material to extreme temperatures, causing it to emit vast amounts of radiation. This process not only influences the growth and evolution of black holes but also {makes impacts on} their surroundings, contributing to phenomena such as {the emergence of }relativistic jets and energetic outflows. The study of black hole accretion is thus essential for understanding both the {astrophysics} of black holes and their broader astrophysical environments.{In a simplified set up, spherically symmetric accretion onto astrophysical black holes has also been studied. In Bondi's original treatment} the pressure and the density of the gas was always connected by the adiabatic equation of state of an ideal gas:
\begin{equation}
    \frac{p}{p_{\infty}} = \left(\frac{\rho} {\rho_{\infty}}\right)^{\gamma},
    \label{bondi_eos}
\end{equation}
$\gamma$ being the ratio of the specific heats {at the constant pressure and the constant volume} and $1 \leq \gamma \leq 5/3$. Bondi's analysis is based upon two basic equations. If $r$ be the radial coordinate and $v$ be the inward velocity of the gas, {stationary integral solution of the} continuity equation is given by:
\begin{equation}
    4 \pi r^2 v \rho = \rm{constant}.
    \label{bondi_cont}
\end{equation}
In steady state, the accreting fluid also satisfies the stationary integral solution of the Bernoulli's equation:
\begin{equation}
    \frac{v^2}{2}+ \int^p_{p_{\infty}}\frac{dp}{\rho} - \frac{GM}{r} = \rm{constant} (= 0).
    \label{bondi_bern}
\end{equation}
The right-hand side of equation (\ref{bondi_bern}) vanishes due to the boundary conditions \cite{bondi1952spherically}, and the second term on the left-hand side can be integrated using the equation of state (eq. (\ref{bondi_eos})). Bondi investigated the steady-state solution through an algebraic analysis of eq. (\ref{bondi_cont}) and eq. (\ref{bondi_bern}). Such analysis was facilitated by introducing a variable, the Mach number, which is the ratio of the dynamical velocity (\(v\)) of the gas to the local sound speed (\(c_s = \sqrt{\frac{\gamma p}{\rho}}\)).

Through his algebraic approach, Bondi identified various types of accretion solutions, including those that {make} transition from subsonic to supersonic regimes (or vice versa), referred to as transonic solutions hereafter. Bondi's work constitutes the simplest model of accretion {within the framework of the} Newtonian gravity framework, fully exploiting the spherical symmetry and by neglecting any {rotational energy component} of the in-falling material. The term Bondi Flow (or Bondi Accretion) is now synonymous with spherical, inviscid, steady accretion in Newtonian gravity.

However, rather than using an algebraic approach, it is often more practical to solve a system of differential equations {to obtain the stationary, transonic accretion solution in the spherically symmetric flow}. Additionally, instead of applying boundary conditions at infinity, it is computationally more efficient to consider boundary conditions at the critical point of the flow, which will be explained in subsequent sections (see \cite{chakrabari_review} for details).
\section{Phase Portrait for the Bondi Flow}
\begin{figure}
    \centering
    \includegraphics[scale = .65]{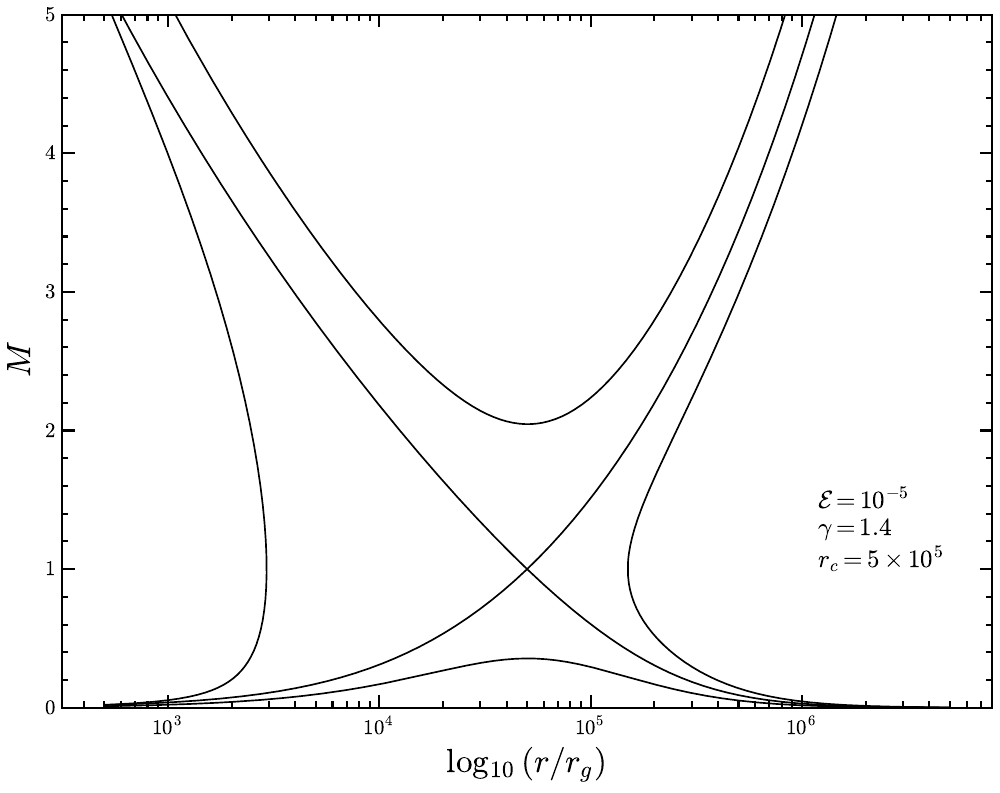}
    \caption{\label{fig:bondi_phase}Phase portrait of the Bondi flow}
\end{figure}

{Stationary integral solutions of the fluid dynamic equations governing the steady state the Bondi flow provide two first integrals of the motion, and it can be shown that the aforementioned two conserved quantities are the specific energy and the mass accretion rate, respectively with their explicit expressions as shown below:}

\begin{equation}
    \mathcal{E} = \frac{1}{2}v^2 + \frac{1}{\gamma -1}c_s^2 - \frac{1}{r} = \frac{1}{\gamma - 1} c_{s_{\infty}}^2,
    \label{bernoulli's}
\end{equation}
\begin{equation}
    \dot{M} = \rho v r^2.
    \label{maccrate}
\end{equation}
where $c_s^2$ represents the local adiabatic sound speed, $c_s^2 = \gamma p / \rho$, utilizing the adiabatic equation of state $p\rho^{-\gamma} = \text{constant}$. Here, $c_{s_{\infty}}$ denotes the sound speed at a considerable distance from the accretor, and $\mathcal{E}$ is identified as the conserved specific energy.
Radial velocity profile of the acreting matter can be obtained by integrating the following equation:
\begin{equation}
    \frac{dv}{dr} = \frac{\frac{1}{r^2} - \frac{2c_s^2}{r}}{\frac{c_s^2}{v} - v} = \frac{N}{D}.
    \label{dvdr}
\end{equation}
Given the continuous nature of physical flow, if \( D \) vanishes at any \( r \), then \( N \) must also vanish at that \( r \). These locations (\( r_c \)) are termed \textit{critical points} or \textit{fixed points}, borrowing the term from the dynamical systems theory . At this point, \( v_c = c_{s_c} \) and \( r_c = \frac{1}{2 c_{s_c}^2} \), {and hence the critical point is actually the sonic point of the flow}. These conditions are known as the \textit{sonic point conditions}. Throughout this manuscript, a subscript `c' denotes values at the \textit{critical points}. The Mach number at the critical point is \( M = 1 \). Substituting the sonic point conditions into equation (\ref{bernoulli's}) yields:
\begin{equation}
    r_c = \frac{5 - 3\gamma}{4(\gamma - 1) \mathcal{E}}.
    \label{rcrit}
\end{equation}
For any \(\gamma\), the  values of all flow variables {at sonic points} are fully determined by the conserved energy, which is in turn determined by the boundary conditions at infinity. Integrating equation (\ref{dvdr}) is simplified with boundary conditions expressed at the flow's critical point. The local sound speed $a(r)$ can be obtained from eq. (\ref{bernoulli's}) once $v_c = v(r_c)$ is known. Alternatively, eliminating \( dv/dr \) from equations (\ref{bernoulli's}) and (\ref{maccrate}) provides a differential equation for \( da/dr \) analogous to eq. (\ref{dvdr}):
\begin{equation}
    \frac{dc_{s_c}}{dr} = \frac{(\gamma - 1)\left(\frac{v^2}{r} - \frac{1}{2r^2}\right)}{\left(c_{s_c}-\frac{v^2}{c_{s_c}}\right)}
    \label{dadr}
\end{equation}
This system of two differential equations ((\ref{dvdr}) and (\ref{dadr})) can be numerically integrated to obtain \( v(r) \) and \( c_s(r) \) at various \( r \). A compact representation of this information is the phase portrait, which plots the Mach number \( M(r) = v(r)/c_s(r) \) versus \( r \). The phase portrait of the Bondi flow is shown in Figure \ref{fig:bondi_phase}.
Figure \ref{fig:bondi_phase} illustrates that only two transonic solutions connect the accretor and infinity. The accreting solution, where the Mach number increases as matter approaches the accretor, becomes supersonic in the region \( r < r_c \). The spherical surface at \( r = r_c \) is often termed the \textit{sound horizon} or \textit{acoustic horizon}, as acoustic disturbances generated within this region cannot cross the \( r = r_c \) surface and must move towards the accretor. Thus, the sound horizon acts like an \textit{event horizon} for acoustic perturbations. We will later explore the non-trivial nature of this analogy.
\section{A dynamical systems perspective}
In steady state, any quantitative information about any of the variables of interest must be obtained from the phase portrait of the system. However, eq. (\ref{rcrit}) may not retain its simple form for more general scenario such as axially symmetric accretion. Depending on the system, this equation can be a higher order polynomial or even a transcendental equation with multiple solution. In such scenarios, where multiple critical points (and consequently multiple sound horizons) appear, obtaining the phase portrait at the first place might become challenging. Fortunately, a simple analogy with the autonomous dynamical systems enables us to at least have some qualitative insight about the transonic accretion profiles without drawing the corresponding phase portrait.

One can introduce a dummy variable (say $\tau$) and write eq. (\ref{dvdr}) as:
\begin{eqnarray}
    \frac{dv}{d\tau} &=& u\left(2c_s^2 - \frac{1}{r} \right) = {N} \\ \nonumber
    \frac{dr}{d\tau} &=& r(v^2 - c_s^2) = {D}
    \label{auton}
\end{eqnarray}
Since the critical points (equilibrium points in the terminology of dynamical systems) are fixed by the flow constants, it is simple to use a perturbative scheme around these points to extract information about the phase orbits (or seperatrics) in the neighbourhood of critical points. Thus, introducing $v = v_c + \delta v$, $c_s = c_{s_c} + \delta c_s$ and $r = r_c + \delta r$, where $\delta$ terms are small first order perturbations, one can linearize the system and obtain from eq. (\ref{auton}):

\begin{eqnarray}
    \frac{d(\delta v)}{d\tau} &=& -v_c^2(\gamma - 1)\delta v + \frac{v_c(3 - 2\gamma)}{r_c^2}\delta r \\ \nonumber
    \frac{d(\delta r)}{d\tau} &=& r_c v_c(1 + \gamma)\delta v + 2 (\gamma - 1)v_c^2 \delta r
    \label{auotonomous}
\end{eqnarray}

This kind of parametrization has been explored in fluid dynamics (e.g. see \cite{bohr1993shallow}). In astrophysical context such parametrization was used in \cite{ray2002realizability,afshordi2003geometrically}. The above set of autonomous system can be written in a compact form:
\begin{equation}
 \begin{pmatrix}
\frac{d(\delta v)}{d\tau} \\
\frac{d(\delta r)}{d\tau}\\
\end{pmatrix} = \begin{pmatrix}
-v_c^2(\gamma -1) & \frac{v_c(3-2\gamma)}{r_c^2} \\
r_c v_c(1+\gamma) &  2 (\gamma-1)v_c^2 \\
\end{pmatrix} \begin{pmatrix}
    \delta v \\
    \delta r \\
\end{pmatrix}
\end{equation}    
or symbolically $\mathbf{Y} = A \mathbf{X}$. Using the solution of the type $\delta v \sim exp(\Omega \tau)$ and $\delta r \sim exp(\Omega \tau)$, one can obtain an eigenvalue equation for the above as $A\mathbf{X} = \Omega \mathbf{X}$, and the eigenvalue itself is obtained from the characteristic equation $\det{(A - \Omega I)} = 0$. This leads to a quadratic equation:
\begin{equation}
    \Omega^2 + \mathbf{Tr}A + \Delta = 0,
    \label{omsqeq}
\end{equation}
where $\Delta = \det{A}$. Since, all the components of $A$ are functions of {various flow variables as defined at critical points}, all are completely specified by the flow constants. Once eq. (\ref{omsqeq}) are obtain they also yield the nature of the critical points themselves. Clearly, the nature of the roots of eq. (\ref{omsqeq}) depend on the numerical values of $\Delta$ and the discriminant $D = (\mathbf{Tr}A)^2 - 4\Delta$. Various possible conditions are summarized in table (\ref{tab:1}).
\begin{figure}[ht]
    \centering
    \includegraphics[scale = .65]{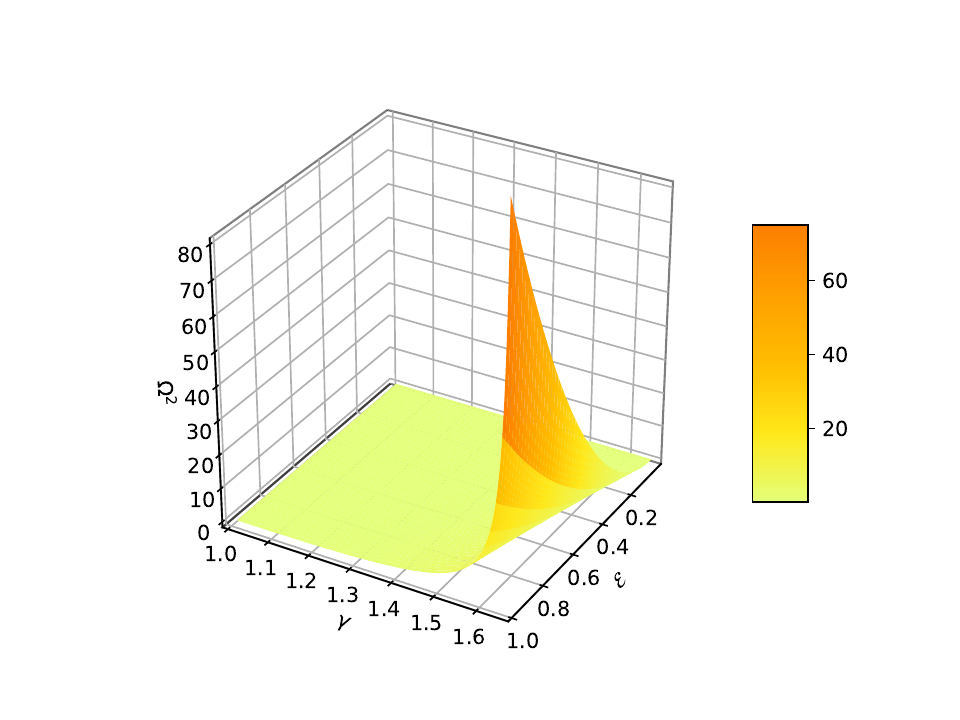}
    \caption{\label{fig:bondi_omega}Phase portrait of the Bondi flow}
\end{figure}
\begin{table}[ht]
\centering
\begin{tabular}{|l|l|l|}
\hline
\textbf{Conditions} & \textbf{Types of roots ($\Omega$)} & \textbf{Nature of critical points} \\ \hline
$\Delta < 0$ & real, opposite sign & Saddle \\ \hline
$\Delta > 0$ and $D > 0$ & real, same sign & Node \\ \hline
$\Delta > 0$ and $D = 0$ & two equal roots & Node \\ \hline
$\Delta > 0$ and $\mathbf{Tr} A = 0$ & purely imaginary roots & Center \\ \hline
$\Delta > 0$ and $D < 0$ & complex roots & Spiral \\ \hline
\end{tabular}
\caption{Types of roots and nature of critical points under different conditions.}
\label{tab:1}
\end{table}
For the present case, there is just one critical point and that is always of saddle type: 
\begin{equation}
    \Omega^2 = \frac{2\mathcal{E}^2(5-3\gamma)}{\left(\frac{1}{\gamma -1} -\frac{3}{2}\right)^2}.
\end{equation}
This is true for any viable parameter value, as is evident from fig.(\ref{fig:bondi_omega}).
The scheme, however, is quite general and in more complicated scenrio its real worth is appreciated \cite{mandal2007critical, chaudhury2006critical}.
\section{Stability of the steady state solution}
In all the discussions above, we relied on the steady state version of the Euler and continuity equations. In {any large-scale astrophysical fluid flow}, such a state may be subjected to external perturbations. A star might pass by the accretor itself and it's gravitational influence might {jeopardize} the steady state configuration at least for some duration. It is thus imperative to check for the stability of the steady state solution against such cases. A standard way to look into such effects is to study the evolution of the {time dependent} perturbations \cite{chaudhury2006critical}. As long as the perturbations do not diverge, the steady state solutions remain meaningful.
The general form of the continuity equation is:
\begin{equation}
    \frac{\partial \rho}{\partial t} + \nabla \cdot (\rho \mathbf{v}) = 0.
    \label{gen_cont}
\end{equation}
For an irrotational flow ($\nabla \times \mathbf{v} = 0 $), satisfies the Euler's equation becomes:
\begin{equation}
    \frac{\partial v}{\partial t} + \frac{1}{2}\nabla(v^2) = -\frac{1}{\rho}\nabla(p) - \nabla\phi,
    \label{gen_euler}
\end{equation}
$\phi$ being the potential of the body force, e.g. gravity. Since, the flow is irrotational, one can introduce a velocity potential $\psi$ such that $\mathbf{v} = \nabla\psi $. Consider some arbitrary but exact stationary solutions of the equations of motion $\left[ \rho_0, v_0, \psi_0\right]$. For a barotropic fluid ($p = p(\rho)$), perturbation in $\rho$ induces a perturbation in $p$ as well around some background value $p_0$. One can then study the linearized perturbations around this background solution by writing: 
\begin{align}
    &\rho(r,t) = \rho_0(r, t) + \epsilon\Tilde{\rho}(r,t)\\ \nonumber
    &p(r, t) = p_0 (r, t) + \epsilon\Tilde{p}(r, t) \\ 
    &\psi(r, t) = \psi_0(r, t) + \epsilon\Tilde{\psi}(r,t), \nonumber
\label{perturbn}
\end{align}
where the overhead \textit{tilde} denotes the perturbed quantity and $0<\epsilon \ll 1$. 
Substituting these into the equation of motions and after some tedious algebra, it is possible to write down the equation that $\Tilde{\psi}(r, t)$ satisfies \cite{visser_analogue}:
\begin{eqnarray}
    & & \frac{\partial}{\partial t}\left(c_s^{-2}\rho_0 \frac{\partial \Tilde{\psi}(r,t)}{\partial t}\right) + \frac{\partial}{\partial t}\left(c_s^{-2}\rho_0 \mathbf{v}_0 \cdot \nabla \Tilde{\psi}(r,t)\right) \nabla \cdot \left(\mathbf{v}_0 \rho_0 c_s^{-2} \frac{\partial \Tilde{\psi}(r,t)}{\partial t}\right) \nonumber\\
    & & + \nabla \cdot \left(c_s^{-2} \mathbf{v}_0 \rho_0 (\mathbf{v}_0 \cdot \nabla \Tilde{\psi}(r,t))\right) - \nabla \cdot \left(\rho_0 \nabla \Tilde{\psi}(r,t)\right) = 0.
    \label{wveqn}
\end{eqnarray}

Eq. (\ref{wveqn}) can be written in a much compact form:
\begin{equation}
    \partial_\mu\left(f^{\mu\nu}\partial_{\nu}\right)\Tilde{\psi}(r, t) = 0,
    \label{gen_wc}
\end{equation}
where $\mu, \nu = 0, 1, 2, 3$ and $f^{\mu\nu}$ is given by:
\begin{equation}
    f^{\mu\nu} = \left(\frac{\rho_0}{c_s^2}\right)\begin{pmatrix}
        1 & v^1 & v^2 & v^3 \\
        v^1 & v^1v^1-a^2& v^1v^2 & v^1v^3\\
        v^2 & v^2v^1 & v^2v^2-^2& v^2v^3\\
        v^3& v^3v^1 & v^3v^2 & v^3v^3 -c_s^2\\
    \end{pmatrix},
\end{equation}
where $(v^1, v^2, v^3)$ are the components of the $3$-velocity $\mathbf{v}$. For the spherically symmetric Bondi flow, of course, the only velocity component available is the radial velocity which we denote simply with $v$. The dummy indices in eq. (\ref{gen_wc}) then runs for $(0, 1)$ only and the $f^{\mu\nu}$ in this case is given by:
\begin{equation}
    f^{\mu\nu} = \left(\frac{\rho_0}{a^2}\right)\begin{pmatrix}
        1  & v \\
        v & (\left|v\right|^2 - a^2) \\.
    \end{pmatrix}
    \label{fmu_bondi}
\end{equation}
An wave equation such as eq. (\ref{gen_wc}) opens up new possibilities to explore which we shall do in the next section. For the moment let us concentrate on the stability of the stationary solution which we sought to investigate.
Consider a trial acoustic wave function for wave equation for the Bondi flow:
\begin{equation}
    \Tilde{\Psi} = g_\omega(r)\exp{(-i \omega t)}
\end{equation}
Then the spatial part $g_\omega(r)$ satisfies
\begin{equation}\label{acstc_wv_eqn}
     -\omega^2 f^{tt}g_\omega - i\omega (f^{tr}\partial_r g_\omega +g_\omega\partial_r f^{rt}+f^{rt}\partial_r g_\omega)+(g_\omega \partial_r f^{rr} +f^{rr} \partial_{rr} g_\omega)=0
\end{equation}
where $\displaystyle f^{tt} = \frac{v_0}{\Psi_0}$, $\displaystyle f^{tr} = f^{rt} =  \frac{v_0^2}{\Psi_0}$ and $\displaystyle f^{rr} = \frac{v_0}{\Psi_0}(v_0^2-c_{s_0}^2)$. The solution can represent a standing wave as well as a travelling wave.
\subsection{Standing Wave Analysis}
\label{sec:stdwv}
If the accretor is {an astrophysical object with a }well defined {physical boundary}, one can assume the perturbations to vanish at the two extremes, far away from the accretor and on its surface. A standing wave solution is useful in such scenarios. However, as a transonic flow becomes supersonic at the sound horizon the only way it can become subsonic again is through a discontinuous shock transition. But a standing wave solution is continuous. So, for the flow to have a standing wave solution it has to be \textit{subsonic throughout}. 
Let the two boundaries be $r_1$ and $r_2 (>r_1)$ such that $g_\omega(r_1) = g_\omega(r_2) = 0$. Now multiplying equation \ref{acstc_wv_eqn} and integrating in the range $r_1<r<r_2$ we get
\begin{equation}\label{omgsq_eqn}
     \omega^2 \int g_\omega^2 f^{tt}dr + i\omega \int \partial_r (g_\omega^2 f^{rt})dr + \int f^{rr}(\partial_r g_\omega)^2dr=0
\end{equation}
After substituting the values of metric elements $f^{\mu\nu}$ we have
\begin{equation}
    \omega^2 =-\frac{\int v_0 (v_0^2-c_{s_0}^2)(\partial_r g_\omega)^2dr}{\int v_0 g_\omega^2 dr}
\end{equation}
As the flow is subsonic everywhere, $v_0 < c_{s_0}$, then $\omega$ is real and we get oscillatory solution for standing wave and the perturbations never diverge.
\subsection{Travelling Wave Analysis}
\label{sec:trwv}
If the accretor happens to be a black-hole, there is no hard surface and any flow that reaches the event horizon is supersonic. In such a scenario the analysis described in sec. (\ref{sec:stdwv}) is not {valied}. The wave in this scenario is a travelling one. The wavelength of such a wave though is small compared to the radius of the accretor. Hence, employing \textit{WKB approximation method} we can approximate the solution in power series of $\omega$
\begin{equation}\label{gw_expan}
    g_\omega(r)= \exp\left[\sum_{n=-1}^{n=\infty} \frac{k_n(r)}{\omega^n}\right]
\end{equation}
Substituting this in equation \ref{omgsq_eqn} and equating the coefficients of $\omega$'s we get the following three equations

\begin{align}
    &-f^{tt}-2if^{tr}\frac{dk_{-1}}{dr}+f^{rr}\left(\frac{dk_{-1}}{dr}\right)^2=0\\
    &-2if^{tr}\frac{dk_0}{dr}-i\frac{df^{tr}}{dr}+\frac{df^{rr}}{dr}\frac{dk_{-1}}{dr}+f^{rr}\left[2\frac{dk_{-1}}{dr}\frac{dk_0}{dr}+\frac{d^2k_{-1}}{dr^2}\right]=0\\
    &-2if^{tr}\frac{dk_1}{dr}+\frac{df^{rr}}{dr}\frac{dk_0}{dr}+f^{rr}\left[2\frac{dk_{-1}}{dr}\frac{dk_1}{dr}+\left(\frac{dk_0}{dr}\right)^2+\frac{d^2k_{0}}{dr^2}\right]=0.
\end{align}
Solving for $\displaystyle\frac{dk_{-1}}{dr}$, $\displaystyle\frac{dk_{0}}{dr}$ and $\displaystyle\frac{dk_{1}}{dr}$ and then integrating using the values of metric elements we obtain:
\begin{align}
    &k_{-1} =  i\int\frac{dr}{v_0 \mp c_{s_0} }\\
    &k_0 = -\frac{1}{2}\ln\left(\frac{v_0c_{s_0}}{\Psi_0}\right)
\end{align}
Thus for $\omega \ll 1$ the first three terms of $g_\omega(r)$ expression (eq. \ref{gw_expan}) are $\sim \omega r, \ln r$ and $1/(\omega r)$ respectively. For large values of $r$:
\begin{align}
    &\omega r \ll \ln r \ll \frac{1}{\omega r}\\
    \implies \hspace{0.3cm}&\omega\mid{k_{-1}}\mid\ll\mid k_0\mid\ll \frac{1}{\omega}\mid k_{1}\mid
\end{align}
Hence, power series in $g_\omega(r)$ does not diverge even as $n$ increases, that is, 
\begin{equation*}
    \omega^{-n}\mid k_n(r)\mid \ll \omega^{-(n+1)}\mid k_{n+1}(r)\mid,
\end{equation*}
and one obtains a travelling wave solution with finite amplitude.

Thus, in either case, i.e., standing wave analysis as described  in sec. (\ref{sec:stdwv}) and  travelling wave analysis as described in sec. (\ref{sec:trwv}), the stability of the steady state solution is ensured.
\section{Analogue spacetime}
It is interesting that the eq. (\ref{fmu_bondi}) looks similar to the equation satisfied by a massless scalar field ($\varphi$) propagating in a curved spacetime:
\begin{equation}
    \partial_{\mu}\left(\sqrt{-g}g^{\mu\nu}\partial_{\nu}\varphi\right) = 0.
    \label{scf}
\end{equation}
A direct comparison of eq. (\ref{scf}) with eq. (\ref{fmu_bondi}) and some algebraic manipulation lead to an expression for the {space-time metric governing the propagation of the linear perturbations embedded within the fluid}:
\begin{equation}
    g_{\mu \nu}= -\frac{1}{c_{s_0}}     \begin{bmatrix}
        v_0^2-c_{s_0}^2 &  -v_0 \\
        -v_0 & 1 
    \end{bmatrix},
    \label{anmetric1}
\end{equation}
and the {corresponding line element (apart from the conformal factor)}:
\begin{equation}
    ds^2 = \frac{1}{c_{s_0}}[(v_0^2 - c_{s_0}^2)dt^2 + dr^2 - 2v_0drdt].
    \label{anle}
\end{equation}
The line element is similar to the Schwarzschild line element in Painlevé-Gullstrand coordinates and has a nonzero curvature. However, since eq. (\ref{fmu_bondi}) holds for acoustic perturbations, the curved spacetime is visible only to these perturbations, while the background spacetime remains Newtonian. The emergence of such an effective metric in the context of an inviscid, irrotational, classical fluid (known as the "acoustic metric") has been recognized since Unruh's pioneering work in 1981 \cite{unruh1981experimental},{and has further been formalized by Visser\cite{visser_analogue}}. Models using this concept are called `analogue gravity' models (for detailed discussion, see \cite{barcelo2011analogue}).
In astrophysical accretion, analogue gravity models are unique, as they may involve both a \textit{gravitational} event horizon and an \textit{acoustic} horizon (if the accretor is a black hole). Once the analogy is established for a given configuration, {certain methodologies as developed in the theory} of General Relativity can be used to study the causal structure of the spacetime. Earlier, a similarity between the sound horizon and a black hole event horizon was noted. It is particularly interesting to establish that the sound horizon is indeed a null horizon in the causal structure of the acoustic spacetime. With a steady-state solution (and a phase portrait such as fig. (\ref{fig:bondi_phase})), we can calculate the acoustic spacetime at every \(r\). Coordinate transformations can then be used to draw the Carter-Penrose diagram, which visualizes the causal structure. The procedure is discussed in the next section.
\subsection{Carter-Penrose diagram for the Bondi flow}
We start by choosing null (sonic) coordinates to write down the line element (\ref{anle}). We note that for null curves (soundlike in case of analogue space-time) $ds^2 =0$, which yields
\begin{equation}
\left(dt-A_{+} (r) dr\right)\left(dt-A_{-} (r) dr\right)=0
\end{equation}
where,
\begin{equation}
A_{\pm} = \frac{-g_{tr} \pm \sqrt{g_{tr}^2-g_{rr}g_{tt}}}{g_{tt}} .
\end{equation}
So instead of co-ordinates $(t,r)$, we choose new co-ordinates to be null co-ordinates $(\chi ,\omega)$ such that
\begin{eqnarray}\label{null_co_ordinates_differential}
d\omega = dt -A_{+}(r)dr\\
d\chi = dt -A_{-}(r)dr
\end{eqnarray}
Using coordinate transformation introduced in (\ref{null_co_ordinates_differential}), the line element (\ref{anle}) can be written as
\begin{equation}
ds^2 = g_{tt} d\chi d\omega
\end{equation}
We now expand $A_- (r)$ and $A_+ (r)$ up to first order of $(r-r_c )$. Thus by expanding $v_0$ near $r_c$ as
\begin{equation}\label{expansion}
v_0 (r) = -v_c  + \left| \frac{dv}{dr}\right|_{r_c }(r-r_c ) + O\left( (r-r_c)^2\right)
\end{equation}
Expanding $c_{s_0}$ in a similar manner we can show that:
\begin{equation}\label{divergent_factor_upto_first_order}
v_0^2-c_{s_{0}}^2 \approx -2 (v_0)_c\left( v_0^{\prime} - c_{s_0}^{\prime}\right) (r-r_c )
\end{equation}
considering up to the first order terms, where $\prime$ denotes derivative with respect to $r$.\\
Now we expand $A_- (r)$ and $A_+ (r)$ up to linear order of $(r-r_c ) $. For that we first note that $g_{tt} \propto (u_0 - c_{eff})^2$ is very small near $r_c$ which implies $|g_{tt}g_{rr}/g_{tr}^2| \ll 1$. Thus we obtain
\begin{eqnarray}
A_+ (r) & = & \frac{ -g_{tr} + g_{tr}\left( 1- \frac{g_{tt}g_{rr}}{g_{tr}^2}\right)^{1/2}}{g_{tt} } \\
& \approx & -\frac{g_{rr}}{2g_{tr}}
\end{eqnarray}
and
\begin{eqnarray}
A_- (r) & = & \frac{ -g_{tr} - g_{tr}\left( 1- \frac{g_{tt}g_{rr}}{g_{tr}^2}\right)^{1/2}}{g_{tt} } \\
& \approx & -\frac{2g_{tr}}{g_{tt}}\\
& = & \frac{1}{\kappa}\frac{1}{r-r_c}
\end{eqnarray}
where
\begin{equation}
\kappa = (v_0^{\prime})_c - (c_{s_0}^{\prime})_c.
\end{equation}
Even though
\begin{equation}
\chi \approx t- \frac{1}{\kappa}\ln|r-r_c | 
\end{equation}
shows a logarithmic divergence at $r = r_c$, the form of $g_{rr}$ and $g_{tr}$ ensure that
\begin{equation}
\omega = t + \int \frac{g_{rr}}{2g_{tr}} dr
\end{equation}
 is regular there.

Thus, near critical points
\begin{equation}\label{divergent_exponent}
e^{-\kappa \chi} \propto e^{-\kappa t}\left| r-r_c \right| \propto  e^{-\kappa t} (v_0^2-c_{s_0}^2)
\end{equation}
In this case, one can compare the acoustic null coordinates with those of the Schwarzschild metric and deduce a coordinate transformation that removes the singularity of the metric element at the critical point. The transformation equations are given by:
\begin{eqnarray}
\left.\begin{aligned}
U(\chi) &= -e^{-\kappa\chi}\\
W(\omega) &= e^{\kappa\omega}
\end{aligned}\right.
\end{eqnarray}
This is analogous to transforming to a coordinate system for the Schwarzschild metric, which behaves regularly at \(r = r_c\). Finally, to compactify the infinite space into a finite patch, we use coordinates \((T, R)\) such that:
\begin{eqnarray}
T= \tan^{-1} (W)+\tan^{-1} (U)\\
R= \tan^{-1} (W)-\tan^{-1} (U).
\end{eqnarray}
\begin{figure}
    \centering
    \includegraphics[scale =.65]{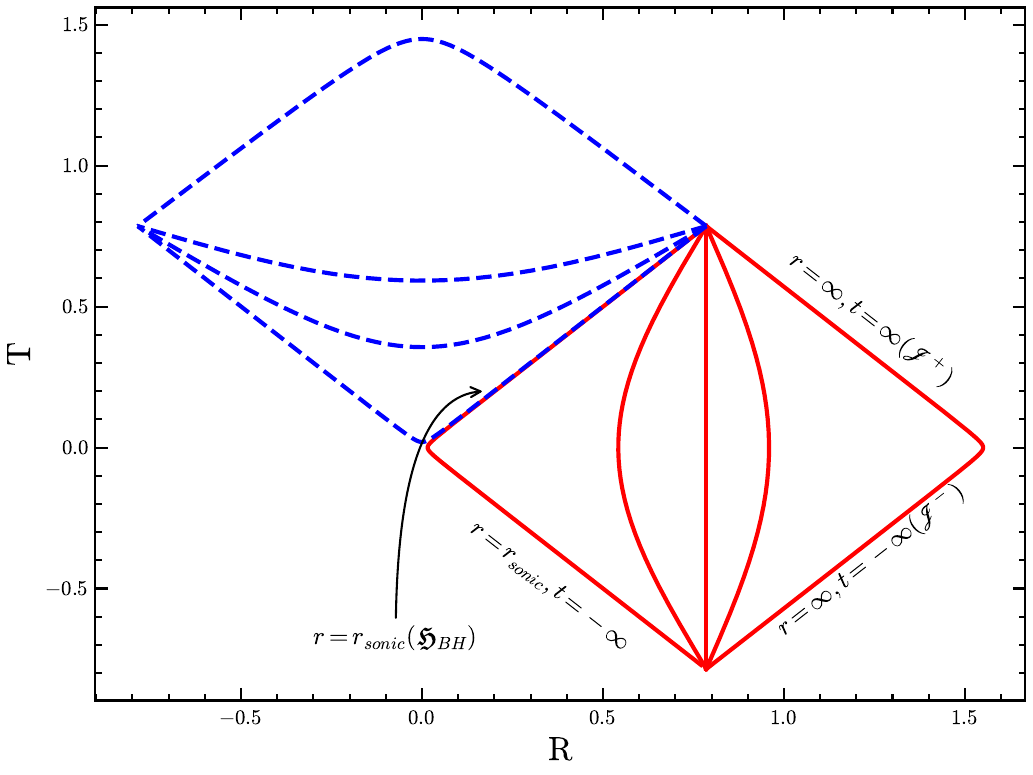}
    \caption{\label{fig:bondi_cp}Carter-Penrose diagram for the analogous spacetime of the Bondi flow.}
\end{figure}
In $(T,R)$ coordinates, lines where $r=\rm{constant}$ can be drawn, allowing the resulting diagram to represent the causal structure of the original spacetime in a compactified region. This diagram, shown in Figure \ref{fig:bondi_cp}, is known as the Carter-Penrose diagram (hereafter referred to as CP). For brevity, we will not delve into the algebra or detailed methodology for drawing the CP diagram. Readers interested in the usefulness of CP diagrams for exploring black-hole spacetimes are referred to Townsend's excellent review \cite{hawking2023large} or \cite{townsend1997black} for more mathematical rigor. The detailed methodology for obtaining CP diagrams in the context of analogue gravity is discussed in \cite{cp_visser} and \cite{maity2022carter}. Here, we will highlight some features of CP, but first, it is helpful to quote a lemma from differential geometry \cite{fre2013gravity}.

\textit{If two metrices $G$ and $g$ on the same manifold $\mathscr{M}$ are conformally related, then the null geodesics with respect to metric $G$ are null geodesics also with respect to the metric $g$ and vice-versa.}
Thus any perturbation propagating with the sound-speed is null-like in the acoustic space-time. One can define the boundary of the mapping $\psi$ of the entire analogue space-time $\mathscr{M}$ as
\begin{equation}
\partial \psi\left( \mathscr{M} \right) = i^0 \bigcup \mathscr{J}^+ \bigcup \mathscr{J}^-
\end{equation}
where
\begin{enumerate}
\item $i_0$, known as \textit{Spatial Infinity} is the endpoint of the $\psi$ image of all space-like curves in ($\mathscr{M}$, $g$).
    
\item $\mathscr{J}^+$, known as \textit{Future Causal Infinity} is the endpoint of the $\psi$ image of all future directed causal curves in ($\mathscr{M}$, $g$).
    
\item $\mathscr{J}^-$, known as \textit{Past Causal Infinity} is the endpoint of the $\psi$ image of all past directed causal curves in ($\mathscr{M}$, $g$).
\end{enumerate}

The sound horizon is immediately identified as a null hypersurface from the fig. (\ref{fig:bondi_cp}). We have already shown that no acoustic perturbation, created within the sound horizon (blue region in the fig. (\ref{fig:bondi_cp}) can cross the sound horizon and escape to a large distance away from the accretor (blue region in the fig. (\ref{fig:bondi_cp}). That along with the fact that the sound horizon is a null horizon in the acoustic space-time shows that sound horizon behaves like an event horizon for acoustic perturbations.
\section{Conclusion}
The procedures discussed in the present article is fairly general and are applicable to any inviscid, irrotational fluid flow. In the context of accretion, Bondi flow is the simplest possible configuration and are discussed merely for a demonstrative purpose. The methodology has been consistently applied to various accretion disk models both in Newtonian and in General relativistic frameworks \cite{chaudhury2006critical, maity2022carter}. {One immediate application of the Carter-Penrose diagram is worth noting. For axially symmetric accretion disks in hydrostatic equilibrium along vertical directions, the \textit{sonic points} and \textit{critical points} do not coincide. In such scenarios the Carter-Penrose diagrams conclusively show that the horizons lie on the \textit{critical points} of the flow and not on the \textit{sonic points} \cite{maity2022carter}.}


\end{document}